\documentclass[pdflatex,sn-apa,oneside]{sn-jnl}
\usepackage{amsmath,amssymb}

\AtBeginDocument{\setlength{\evensidemargin}{\oddsidemargin}}
\usepackage{mathptmx}

\begin{document}

\hypersetup{
  pdftitle={Intrinsic Computational Functionalism: From Observer-Relative Maps to Observer-Independent Structures},
  pdfauthor={Shuqin Ma; Ryota Kanai},
  pdfsubject={Philosophy of mind; consciousness studies},
  pdfkeywords={computational functionalism, consciousness, observer-relativity, intrinsic computation, causal organisation, artificial consciousness}
}

\title[Intrinsic Computational Functionalism]{Intrinsic Computational Functionalism: From Observer-Relative Maps to Observer-Independent Structures}

\author*[1,2]{\fnm{Shuqin} \sur{Ma}}\email{23110160046@m.fudan.edu.cn}\equalcont{Equal contribution.}

\author*[3]{\fnm{Ryota} \sur{Kanai}}\email{kanair@araya.org}\equalcont{Equal contribution.}

\affil*[1]{\orgdiv{School of Philosophy}, \orgname{Fudan University}, \city{Shanghai}, \country{China}}

\affil[2]{\orgname{Sussex Centre for Consciousness Science}, \orgaddress{\orgname{University of Sussex}, \country{United Kingdom}}}

\affil[3]{\orgname{Araya, Inc.}, \orgaddress{\city{Tokyo}, \country{Japan}}}

\abstract{Anti-computational arguments show that externally imposed computational interpretations cannot ground consciousness, but they do not establish that all computational organisations are observer-relative. We develop \textit{intrinsic computational functionalism}: the view that, if consciousness is computationally constituted, it depends on physically realised computational structures the system has in virtue of itself rather than on labels imposed by an external interpreter. Two criteria operationalise this view. (C1) System-intrinsic instantiation: the relevant property must be specifiable without an observer’s labelling, and invariant under structure-preserving relabellings of the system’s variables. (C2) Causal-dynamical organisation under intervention: the property must be grounded in a state-space structure whose variables mutually constrain one another, and whose organisation is exhibited in counterfactual response under intervention. Together these criteria specify what any candidate computational account must satisfy to remain observer-independent, without selecting which intrinsic structures bear on experience. The argumentative core is a three-tier decomposition of identification work: interpreter-relative label selection (tier i), theoretically constrained partition selection (tier ii), and dynamics-internal grain selection (tier iii). We argue that any computational property capable of avoiding the observer-relativity objection must be identified, if at all, through tier (iii) dynamics-internal grain selection, conditional on empirically disciplined tier (ii) choices. Syntax-is-not-semantics arguments, mapmaker arguments, and the observer-relativity component of biological-naturalist objections succeed against views that locate the consciousness-relevant property at tier (i); once the tiers are distinguished, intrinsic computational functionalism survives.}


\maketitle

\noindent\textbf{ORCID iDs:} Shuqin Ma \href{https://orcid.org/0009-0008-7375-7297}{0009-0008-7375-7297}; Ryota Kanai \href{https://orcid.org/0000-0002-0186-2687}{0000-0002-0186-2687}.

\section{Introduction}

Artificial systems now produce fluent self-descriptions and behave in ways that invite attributions of consciousness. This raises the question whether their internal organisation is of a kind that could constitute consciousness at all. Within computational and functionalist approaches, one strategy is to identify a candidate property whose presence in a system would be relevant to consciousness, and evaluate whether a given system possesses that property (\citealp{butlin2023consciousness}; \citealp{butlin2025identifying}). But whether any computational property can play this role has been contested on several fronts.

A recurring family of arguments holds that no candidate property of a computational kind can do the work consciousness requires. Syntax-is-not-semantics arguments hold that pure rule-following on symbols yields no understanding, and the later metaphysical variant holds that syntax is not intrinsic to physics, since the same physical states admit incompatible syntactic readings \citep{searle1980minds, searle1992rediscovery}. Biological naturalism ties consciousness to the brain’s specific biological causal powers rather than to abstract computation \citep{searle1992rediscovery}, with related biological and enactive approaches locating it in metabolic, self-maintaining, and organismic activity \citep{seth2025conscious, thompson2007mind}. Triviality arguments hold that sufficiently complex physical systems can be mapped onto a wide range of computations under some labelling \citep{putnam1988representation}, and mapmaker arguments hold that computational symbols do not pre-exist in physics but are alphabetised into existence by a cognitive agent who maps continuous dynamics onto a finite predefined vocabulary \citep{lerchner2026abstraction}. The objections do not share a single argument, and some raise challenges that go beyond observer-relativity. They converge on the worry that computation, on familiar specifications, omits something that consciousness requires of any candidate property, whether interpreter-independence, biological causal power, organismic world-involvement, or labelling-invariance.

These objections expose a genuine problem for interpretation-based computational functionalism that treats implementation as mere formal matching under freely chosen interpretations. A computational functionalism that requires only a purely formal structure to be realised under any interpretation is a triviality position, since any physical system can then be mapped onto any computation if the mapping is chosen freely, making consciousness observer-relative. The objection succeeds against this weak form of computational functionalism, but it does not by itself refute the position in which consciousness depends on physically realised computational organisation rather than on interpretive labelling.

The central argument is that anti-computational objections conflate two things: the observer-relativity of computational interpretation and the observer-relativity of computational implementation. The three-tier decomposition developed in Section 7 separates the two, and the criteria (C1) and (C2) operationalise the distinction. We call this position \textit{intrinsic computational functionalism}: the view that the consciousness-relevant property, if computational, must hold in virtue of the system’s own organisation, independently of any observer’s labelling. Two criteria operationalise it (stated and defended in Section 5): (C1) system-intrinsic instantiation, and (C2) causal-dynamical organisation under intervention. Both must be met by any candidate consciousness-relevant computational organisation; neither selects the consciousness-relevant subset on its own. Further substantive conditions, including recurrent self-modelling, global availability, and non-trivial information closure, must be defended on their own terms. We diagnose anti-computational arguments as equivocating between tier (i) labelling-dependence and tier (iii) implementation-dependence. The tiers are levels of identification work: arbitrary symbol assignment at tier (i), theoretically disciplined partition choice at tier (ii), and dynamics-internal grain selection at tier (iii). The claim that causal organisation matters is already shared by mechanistic individuation \citep{piccinini2015physical}, integrated information theory (IIT; \citealp{albantakis2023integrated}), and the universality requirement \citep{kanai2024toward}; the tier decomposition is what is new. While \citet{lerchner2026abstraction} provides the most explicit recent articulation of the mapmaker challenge, the tier framework developed here is independent of any single source and applies to the family of anti-computational arguments as a whole.

Section 2 collects the anti-computational family and identifies what each member correctly establishes. Section 3 distinguishes computational interpretation from computational implementation. Section 4 explains why naive computational functionalism is insufficient and what any successor must rule out, before Section 5 states (C1) and (C2). Section 6 traces the convergent traditions sharing the intrinsic turn. Section 7 separates three tiers of identification work on which the family of anti-computational challenges can be tested. Section 8 applies the criteria to candidate architectures, working through a Hopfield attractor network, a lookup-table system, a standard transformer chatbot, and Beautiful Loop, a recent active-inference theory of consciousness. Section 9 considers how partition selection is empirically constrained. Section 10 draws cautious implications for current artificial systems, and Section 11 outlines the unified research orientation.

The explanatory gap \citep{levine1983materialism} and the hard problem \citep{chalmers1995facing} are bracketed throughout. The argument assumes only that computational structure is a candidate for consciousness-relevance, without claiming that any particular system is conscious. The ‘consciousness-relevant’ qualifier is theory-conditional throughout: it asks whether a property is one a substantive consciousness theory could pick out as constitutive, while leaving open which theory is right. Readers who reserve ‘computation’ for externally interpreted symbol manipulation may regard intrinsic computational functionalism as a related but distinct position rather than a version of computational functionalism.

\section{The anti-computational challenge}

This section surveys the family of anti-computational objections, grants what they establish at the level of imposed labels, and identifies the further premise their generalisations require.

The challenge takes several forms. The Chinese-room argument holds that pure rule-following manipulation of symbols yields no understanding \citep{searle1980minds}, and the later metaphysical version holds that syntax is not intrinsic to physics \citep{searle1992rediscovery}. Searle’s biological naturalism holds that consciousness depends on the biological causal powers of the brain rather than on abstract computation \citep{searle1992rediscovery}; related biological and enactive approaches emphasise metabolic, self-maintaining, and organismic activity as central \citep{seth2025conscious, thompson2007mind}. Triviality arguments hold that every ordinary open physical system implements every finite-state automaton under a suitable state-mapping \citep{putnam1988representation}, trivialising computational implementation (see also \citealp{sprevak2018triviality}). A recent mapmaker formulation makes the worry concrete: a single voltage trajectory admits multiple incompatible labellings with no fact of the matter privileging one, since it can be read as Beethoven’s 5th, its retrograde inversion, or stock-market prices, depending on which labelling is imposed \citep{lerchner2026abstraction}.

Across the objections that target computational interpretation, one point should be granted: externally assigned symbols are observer-relative. This is the labelling layer corresponding to tier (i) of the three-tier decomposition developed in Section 7. Because such labels can be imposed in multiple incompatible ways without altering the underlying dynamics, any position that locates the consciousness-relevant property at this level fails. So does any position that identifies computation with formal mapping onto a transition table independent of mechanism, since consciousness requires more than formal mapping even if a theory of computation does not. This diagnosis is continuous with Piccinini’s criticism of simple mapping accounts of concrete computation. On such accounts, arbitrary mappings from physical states to computational states make it too easy to attribute computations to physical systems, leading towards pancomputationalism and undermining the objectivity of computational implementation \citep{piccinini2015physical}.

What is disputed is the move from the local thesis to a stronger generalisation. The mapmaker argument can be read in two ways. One reading concludes that consciousness cannot be computationally constituted at any level, since any identification of a physical state as a ‘readout’ or ‘hidden state’ has already imported an interpretive labelling. \citet{lerchner2026abstraction} draws a stronger conclusion: computation presupposes consciousness rather than constituting it. Trivialist readings conclude that computational implementation is interpreter-relative without qualification. Each of these generalisations (mapmaker, trivialist) requires a further premise that the local thesis alone does not establish.

A state in a computational mechanism is identified by the functional-causal role it plays within the mechanism, where an observer’s label has no individuating force \citep{piccinini2015physical}. Computational grain can also be specified through dynamics-internal measures, such as effective information, which quantifies how informatively a system’s intervention-conditional transitions distinguish successor states \citep{hoel2017map}, or integrated information theory’s exclusion postulate, which selects the single grain at which cause-effect structure is maximal \citep{albantakis2023integrated}. On neither approach does computation reduce to externally assigned symbol manipulation, and arguments targeting only the narrow conception leave the alternatives in place.

Any scientific account of a system requires choices of variables and boundaries, and that cost is granted. The question is whether the property so identified exists only because the description was imposed, or whether it would be there regardless. Externally assigned symbols exist only under such imposition, since the same voltage trajectory can be read under incompatible labellings with no fact of the matter privileging one reading. But intervention-invariant causal organisation persists across such relabellings: an observer provides the description, but the causal organisation the description identifies is not brought into existence by the act of describing it. Whether this distinction between description-imposed and intervention-invariant structure can be sustained under scrutiny is the question Section 7 addresses.

\section{Interpretation versus implementation}

We distinguish computational interpretation from computational implementation, and everything that follows turns on this distinction. Computational interpretation consists in the labelling work an observer performs: a voltage may be interpreted as ‘1’, a neural response as ‘red’, a hidden vector as ‘cat’, a dynamical trajectory as ‘decision formation’. Such interpretations depend on explanatory aims, measurement conventions, and semantic practices. Computational implementation, by contrast, is a fact about the system’s own organisation: a system implements a computation when its physical organisation realises a pattern of causal state transitions in which its mechanisms make systematic differences to later states under relevant interventions and perturbations, rather than merely passing through a sequence that can be retrospectively mapped onto a computation.

A transistor brings out the distinction. The semantic label ‘1’ is conventional, and the threshold at which a voltage counts as ‘high’ or ‘low’ is specified within an engineering context. But once that interface is fixed, whether the voltage range opens or closes a gate, drives another transistor, stores a memory state, or changes the future behaviour of the circuit is fixed by the device. The labelling does not make the gate switch. A hidden activation in a neural network may likewise be given different semantic interpretations by different observers while its role in constraining downstream activations and actions is part of the system’s physical organisation.

Crucially, the observer-relativity of semantic labels does not imply the observer-relativity of causal structure. A scientific description involves theoretical choices, but those choices do not create the causal organisation they identify. This also addresses the worry that a single physical trajectory can support many computational interpretations. An isolated trajectory is indeed underdetermined, but real computational mechanisms are organised spaces of possible trajectories. This is the move that grounds counterfactual-structure-based responses to Putnam-style triviality \citep{chalmers1996brock}, and that organisation constrains which computational descriptions are legitimate. The distinction is familiar from mechanistic explanation: a heart can be described as a pump, an endocrine organ, an electrophysiological oscillator, or a developmental system. The plurality of these descriptions does not make it observer-created. Likewise, a neural or artificial system may admit many levels of computational description without its causal organisation being merely interpretive.

A further worry is that computational individuation may depend on representational or environmental relations rather than internal causal structure alone \citep{sprevak2010computation}. Two kinds of external dependence need to be distinguished. Causal-environmental relations to natural kinds are part of the system’s intervention-invariant structure and pose no problem for the present account. Dependence on an interpreter’s stipulation is what the mapmaker objection correctly targets and what intrinsic computational functionalism excludes. The present argument concerns only whether observer-imposed labelling is necessary for computational individuation; it does not settle whether environmental or teleological relations also play a role.

The encoding-decoding tradition presses a further worry. On one influential definition, computation is the use of a physical system to model a target abstract entity, mediated by an encoding-decoding loop relative to a computational entity \citep{horsman2014does}; see also \citet{worden2025computers}. These views are broader, since they treat any encoding-decoding loop as constitutively interpretive. The disagreement is whether the modelling work involved in scientific identification is \textit{constitutively} interpretive or \textit{theoretically constrained}. On the view defended here, partition and intervention-space choices are answerable to predictions and to the system’s causal organisation under intervention. The key disanalogy is that an encoder’s choice of mapping leaves the substrate’s intervention-conditional behaviour intact, whereas the tier (ii) partition and intervention-space choices are answerable to whether the predicted dependencies actually emerge under perturbation. Whether the vector representations of artificial systems are grounded, and in what sense, is itself debated \citep{coelhomollomilliere2026}. The dispute is genuine and is not resolved here. What intrinsic computational functionalism requires is that the encoding-decoding conception of computation not be treated as the only available conception. Whether it is correct is a substantive question the present argument does not need to settle.

\section{Why formal equivalence is insufficient}

Section 3 addressed the worry that computational descriptions are observer-imposed. A further worry remains: even granting that some computational organisation is not merely interpretive, formal transition structure alone may be too weak to ground consciousness.

Two senses of ‘abstraction’ need separating before the critique. Cognitive science routinely abstracts from molecular and electrophysiological detail when it treats neurons as nonlinear functions or networks as recurrent dynamical systems, exposing organisational properties that hold across many physical realisations. The criteria (C1) and (C2) are abstractions of this kind, and they remain answerable to the intervention-conditional causal structure of the underlying systems. The problem is not abstraction as such, but abstraction detached from the causal organisation that makes the abstraction true of the system.

Naive computational functionalism claims that sharing the same formal transition structure is sufficient for sharing the same mental properties. The canonical form is early machine-state functionalism \citep{putnam1967psychological}, in which a creature’s psychological state is a state of its Turing-machine description, with mental kinds individuated by their position in the machine table independently of physical realisation. The view does not ignore internal states, but it individuates them by abstract role alone, without requiring that the causal organisation realising the transitions be preserved.

The view neglects internal mechanisms, since a lookup table and a recurrent neural system can produce the same outputs across a range of inputs while differing radically in internal organisation, and if consciousness depends on internal structure, input-output equivalence will not suffice. It neglects temporal and dynamical properties, since consciousness may depend on ongoing self-conditioning dynamics that static functional mappings cannot capture. It neglects causal grain. A system may formally simulate a causal process at one level while failing to realise the causal organisation that matters at another. The question of which grain is the right one cannot be settled by the formal transition structure alone, since many grains are compatible with the same abstract description. An independent criterion is needed to select the grain at which the system’s own causal organisation does its work: this is the role tier (iii) will play in Section 7.

Chalmers’s principle of organizational invariance avoids some of these problems. It holds that any two systems with the same fine-grained functional organisation will have qualitatively identical experiences \citep[Ch. 7]{chalmers1996aconscious}. This is stronger than machine-table equivalence, since it requires the causal topology of internal states to be preserved. The present view is not a rejection of organizational invariance, but a further constraint: it asks that the relevant organisation be identified through dynamics-internal criteria and tested under intervention, rather than merely postulated at a level of abstraction. The gap between organizational invariance and intrinsic computational functionalism is the space the two criteria below are designed to fill.

The relevant computational organisation must be intrinsic, mechanistically realised, dynamically structured, and counterfactually individuated; it must be a property of the system itself rather than of an external description. The two criteria below operationalise this revision.

\section{Two criteria for intrinsic computational organisation}

Intrinsic computational functionalism can be operationalised by two criteria. Each is a necessary condition for a computational organisation to be a candidate for consciousness-relevance, though neither individually nor jointly is sufficient for consciousness. They identify the minimum form that any consciousness-relevant computational property must have if it is to be a candidate constituent of consciousness rather than an artefact of interpretation. They operate at the level of physical dynamics, characterised in terms of cause-effect coupling, internal organisation, and response under intervention. What they require is observer-independence at that level; determining which intrinsic structures are consciousness-relevant is a further question.

\textbf{(C1) System-intrinsic instantiation.} We treat a property as system-intrinsic when its specification does not require an external interpreter’s vocabulary, so that the property’s holding or not holding is a fact about the system characterisable without reference to anyone’s labelling. The causal pattern of pressure generation and fluid displacement that a heart exhibits is system-intrinsic in this sense: it is fixed by what the organ does, not by anyone calling it a pump. A coin’s role as ‘heads’ or ‘tails’ under a betting convention is not: nothing about the coin itself fixes which side counts as ‘heads’. A triangle sharpens the point. Consider a triangle in a coordinate space with $x$- and $y$-axes. The triangle’s mathematical properties (its angles, side ratios, and congruence relations) do not change under swapping $x$ and $y$, rotating the frame, or reflecting it. The same mathematical object persists across all such relabellings of the coordinates. The analogy illustrates relabelling-invariance, not a full theory of intrinsicness. (C1) requires the same of any consciousness-relevant computational property: it must hold in virtue of the system’s intrinsic structure, in the way triangularity holds in virtue of the figure, not in virtue of which coordinate frame an observer imposes. The criterion is methodological. (C1) is not intended as a sufficient metaphysical account of intrinsicness; it is a filter against label-dependent properties, conditional on empirically disciplined tier (ii) partitioning. It does not by itself fix what \textit{kinds} of intrinsic property bear on consciousness, but only what counts as non-label-dependent for the purposes of the present framework.

(C1) has a precise operational form. A property is system-intrinsic when its truth value is invariant under structure-preserving relabellings of the system’s variables. A relabelling is an operation that reassigns names to the system’s variables without altering the dynamics: it permutes indices, renames states, or changes descriptive vocabulary, but leaves the transition structure and intervention-conditional behaviour intact.

Arbitrary remappings of names to variables are observer-imposed, and the consciousness-relevant property cannot live at the level of properties that change under such remappings. Structure-preserving relabellings, by contrast, leave intervention-conditional dynamics intact, and properties invariant under them are facts about the system. The test makes (C1) testable on any fully described dynamical system: for a candidate property $P$ and system $S$, ask whether $P(S)$ is fixed once $S$’s intervention-conditional structure is fixed, independently of how the variables are named or interpretively re-described, and conditional on an empirically constrained partition and intervention space.

\textbf{(C2) Causal-dynamical organisation under intervention.} We use ‘causal-dynamical organisation’ in the physicalist substrate-realised sense: the system’s own intervention-conditional state-space structure, where its variables mutually constrain one another, so that changing one variable alters the future evolution of others. It does not refer either to abstract variable-level intervention relations divorced from substrate or to the componential decomposition of higher-level phenomena in classical mechanistic explanation. Mechanistic individuation \citep{piccinini2015physical}, effective-information maximisation \citep{hoel2017map}, and integrated information theory’s intrinsic cause-effect structure \citep{albantakis2023integrated} share, in different ways, the commitment that the relevant structure is identified through intervention-conditional organisation rather than imposed labelling. They differ in how they specify the privileged grain and in their further metaphysical commitments. (C2) draws on what they share without adjudicating between them.

(C2) has two dimensions, internal organisation and intervention profile, tested separately below. A property satisfies (C2) when it is grounded in an organised state-space structure in which the system’s variables mutually constrain one another, and when the form of that organisation is exhibited in the system’s counterfactual response under intervention on those variables. The first excludes isolated trajectories and lookup tables; the second excludes recorded traces and surface simulations. The requirement that implementation be fixed by counterfactual transition structure rather than actual trajectory is continuous with the counterfactual-conditional framework developed in response to Putnam-style triviality \citep{chalmers1996brock}.

A property satisfies (C2) only when both dimensions are met. The first, an internal-organisation test, asks whether the property can be specified by pointing to any single trajectory through state space, or whether its specification requires the organised structure itself, with forms of internal richness including input sensitivity, recurrent feedback, and error correction. The second, a counterfactual-structure test, asks whether interventions on the variables produce characteristic effect distributions that differ across distinct underlying mechanisms, even when the systems’ surface outputs match under their actual inputs; the relevant intervention-conditional structure can be specified through dynamics-internal measures such as effective information \citep{hoel2017map}, which select the grain at which the system’s counterfactual response is most informative.

Cognitive-neuroscience interventions such as transcranial magnetic stimulation, optogenetic stimulation, and focal lesions are concrete examples of this kind of test, since each asks whether a neural population’s causal contribution under perturbation differs from its actual-input trajectory.

The relevant notion of intervention is dynamics-changing, not label-changing: an intervention sets a variable’s value (or perturbs it within an admissible range) and the system evolves from that new initial condition under its own dynamics \citep{woodward2003making}, so relabelling a variable does not constitute an intervention in this sense. The distinction matters because the family of objections rehearsed in Section 2 trades on conflating the two operations: a relabelling is observer-imposed and trajectory-preserving, whereas an intervention is dynamics-modifying and observer-independent in its effects. The criterion is structural; (C2) requires the property to supervene on organised state-space structure that no relabelling of a single trajectory can preserve, so the question ‘how much integration is enough’ cannot be answered at the level of single-trajectory analysis.

The two criteria filter different failure modes. (C1) rules out properties that vary across re-descriptions of the same dynamics; (C2) rules out properties that appear identical at the surface but come apart under dynamics-changing perturbation.

Two systems with matching surface input-output behaviour illustrate the point. A lookup table that reproduces a target system’s input-output mapping fails the internal-organisation test by representing the property as a tabulated trajectory, not an organised structure of mutually constraining variables. A recorded-trace implementation that replays an integrated system’s surface dynamics fails the counterfactual-structure test, since interventions on its intermediate states leave the playback degenerate where the original mechanism would generate characteristic effect distributions. A candidate mechanism passes both. Given a specified dynamics-internal criterion and intervention space, the relevant grain is fixed by the system’s transition structure rather than by the theorist’s semantic labelling.

A bimetallic strip’s bending in response to ambient temperature is fixed by its own thermal-mechanical structure, so it satisfies (C1); perturbing its temperature produces a structurally robust, deterministic bend response, so it passes (C2)’s counterfactual-structure side trivially. It fails (C2)’s internal-organisation side: at the simplified explanatory grain considered here, the system lives on a one-dimensional state space without mutual constraint among multiple variables. (C1) and the counterfactual side of (C2) can be jointly satisfied while the internal-organisation side fails. A system whose actual dynamics are integrated but whose response to intervention is collapsed (by stored values, external clamping, or stipulated protocol) fails (C2) on the counterfactual-structure side while potentially looking integrated on the surface. Many paradigmatic biological control systems satisfy both sides of (C2) at the relevant explanatory grain: their organisation consists of mutually constraining components, and it exhibits characteristic counterfactual responses to intervention on those components.

What (C2) requires is a structural-conditional counterfactual organisation: intervention-conditional relations among the system’s variables, abstracted from whether those interventions also produce the further metaphysical causal effects (e.g., chemical change) the simulated kind would have. Under interventions on the system’s variables, certain output patterns change in characteristic ways. The criterion does not by itself demonstrate that the property does physical-causal work in the demanding metaphysical sense, since a simulation can in principle satisfy structural-conditional counterfactuals without the simulated property having the metaphysical causal efficacy of the simulated kind. The substrate-based challenge raised by biological naturalism (Section 2) is therefore not settled by these criteria; it is addressed separately in Section 10.

Necessity for each criterion can be argued directly. Drop (C1), and a property counts as obtaining only under an external interpreter’s labelling, which is observer-relative; the property then sits at the layer where the family of anti-computational objections bites. Drop (C2), and input-output equivalence settles computational identity; a lookup table that exhaustively maps any input to any output would be computationally identical to any system with matching surface behaviour, regardless of internal structure. A system that produces every conversational response by table-lookup matches a thinking system’s surface behaviour without sharing any of its internal mechanisms, and most theorists hold the two are computationally distinct. This consequence is rejected even on mechanistic grounds \citep{piccinini2015physical}, independently of any commitment to the simulation-instantiation contrast biological naturalists press.

The criteria are deliberately minimal. They exclude observer-relative properties and lookup-table simulacra, but they do not narrow candidates to the consciousness-relevant subset; that narrowing requires further substantive conditions, including global availability, recurrent self-modelling, non-trivial information closure, and temporal continuity at relevant scales \citep[on temporal continuity, see][]{kanai2025stream}, each to be defended on its own terms. (C1) and (C2) specify what intrinsic computational organisation amounts to without committing to a specific theory of consciousness.

\section{Convergent traditions}

Three research programmes converge on the requirement that consciousness-relevant computational structure must be specifiable without recourse to an external interpreter, though they disagree on how the relevant structure is identified. The convergence is methodological, not doctrinal; intrinsicness is construed variously as intrinsic cause-effect power, as mechanistic-functional individuation, and as a meta-theoretical requirement on the form that theories of consciousness must take.

The closest philosophical precedent is Chalmers’s principle of organizational invariance: any two systems with the same fine-grained functional organisation will have qualitatively identical experiences \citep[Ch. 7]{chalmers1996aconscious}. This principle already captures the spirit of (C1), since it requires preservation of causal-functional topology, where labels can be reassigned without altering the property. The present criteria add a further requirement: the relevant organisation must be identified through dynamics-internal criteria and tested under intervention, not merely postulated at a level of abstraction. The shared commitment is that the consciousness-relevant property, if computational, depends on the system’s realised organisation, independently of any imposed interpretation.

On the mechanistic-functionalist line, Piccinini’s mechanistic account of concrete computation is the most relevant precedent. Computation, on this account, is not individuated by externally assigned semantic content, but by the organisation of a functional mechanism that manipulates medium-independent vehicles according to rules. Components of a computational mechanism are identified by their causal contributions within that organised mechanism; an observer’s semantic assignment plays no constitutive role \citep[Ch. 7]{piccinini2015physical}. The present view takes from this tradition the idea that computational organisation can be physically realised and mechanistically individuated independently of any interpretation. The present criteria add a requirement Piccinini does not emphasise: the privileged grain should be selected by dynamics-internal measures; on the present account, the theorist’s explanatory interests cannot fix the grain by themselves.

The intrinsic-existence orientation of integrated information theory \citep{albantakis2023integrated} is often treated as an alternative to functionalism, but in the present setting it shares the intrinsic-turn orientation we defend: it rejects observer-imposed structure as a ground and locates the consciousness-relevant property in what the system has by virtue of its own organisation. The relevant question is what causal powers the system has upon itself (what its variables do when intervened upon, or in IIT’s technical sense, the system considered as cause and effect upon itself), and consciousness is identified with an integrated cause-effect structure specified at this intrinsic level. What we draw on is this orientation, not IIT’s specific formalism, and we do not commit to its further metaphysical claims.

The previous two traditions offer substantive accounts of computation or consciousness. A parallel constraint operates at the meta-theoretical level: any theory of consciousness must be able to determine, for any fully described dynamical system, whether it is conscious, and its determinant must be intrinsic to the system, not fixed by external interpretation \citep{kanai2024toward}. The requirement bears on the \textit{form} of the criterion, not on the physical vocabulary in which dynamics is described. What the requirement demands is that the question whether a given system is conscious be answerable without recourse to an interpreter.

The programmes may ultimately disagree. Different dynamics-internal grain criteria can select different grains in the same system, and the universality requirement is a constraint on the form of any candidate theory, not a substantive theory in its own right. The disagreements remain open. What the programmes share is enough to constitute a research orientation: interpreter-imposed structure is rejected as a ground, physical realisation is required, and the focus is internal causal organisation. The present paper adds to this orientation a specific diagnostic: anti-computational arguments succeed against properties identified at the level of imposed labels, but do not generalise to properties identified through dynamics-internal causal organisation. The three-tier analysis developed in the next section makes this diagnostic precise.

\section{Three tiers of identification work}

This section separates three operations bundled together in anti-computational arguments and shows that the objections succeed against only one of them. We distinguish interpreter-relative label selection, theoretically constrained partition selection, and dynamics-internal grain selection. Pressing the objection beyond label selection requires a further premise that, we argue, cannot be sustained.

The crucial question is whether identifying a system’s components (its individuated parts), grain (the level of resolution at which states are specified), or state-space partition (the carving of possible states into the variables under consideration) is itself mapmaker work. If it is uniformly so, the position collapses, since the dynamics-internal carving criteria it appeals to would themselves be products of interpretation.\footnote{A recent statement puts the worry plainly: as soon as a physical state is identified as a ‘readout’ or ‘hidden state’, it has already been labelled by a mapmaker \citep[Section 3.2]{lerchner2026abstraction}.}

A further point is that identifying a system’s computational structure involves several distinct operations. One may assign symbolic labels to a physical trajectory, choose a boundary and state-space partition, or select the grain at which the system’s causal-dynamical organisation is analysed. These operations are not observer-dependent in the same way. We therefore distinguish three tiers of identification work. The objection succeeds straightforwardly at the first tier; its extension to the second and third requires a further argument that has not been supplied.

The numbering reflects increasing observer-independence rather than operational sequence: in practice, partition and intervention space at tier (ii) are typically fixed first; dynamics-internal grain at tier (iii) is then determined by the system’s intervention-conditional transitions; and labels at tier (i) can be applied at any stage without affecting the analysis.

\textbf{Tier (i): interpreter-relative label selection.} A physical trajectory is read under a particular labelling, so that the same voltage sequence may be called Beethoven’s 5th, its retrograde inversion, or stock-market prices. The labellings are arbitrary, the system’s behaviour is invariant under relabellings that preserve the causal topology, and no fact about the trajectory privileges one labelling over another. This is where the symbol-assignment objection bites in its strongest form, and we grant it in full: arbitrary symbol-assignment is observer-relative, and any view locating the consciousness-relevant property at this layer fails.

\textbf{Tier (ii): theoretically constrained partition and intervention-space selection.} Any scientific account of a physical system requires three choices: a system boundary, a state-space partition, and an intervention space. We refer to this bundle as the \textit{tier (ii) partition}. These three choices have different degrees of theoretical constraint; we bundle them because each is theory-driven rather than stipulative, and each supports empirically distinguishable commitments. The term covers all three components wherever it appears below. Whether the system is identified as ‘this neural population’ or ‘this brain area’, whether the state space is partitioned by firing rates or membrane potentials, and whether the intervention space includes optogenetic stimulation or only natural perturbations are choices that condition every subsequent claim. The cost is real and is shared by every scientific account. Tier (ii) selection is theory-constrained, not stipulative: it answers to predictions, to fit with observed phenomena, and to the system’s causal organisation under intervention. A partition can be empirically wrong if interventions defined over it fail to reveal the causal dependencies it predicts, or if it systematically obscures stable relations captured by a competing partition, and competing partitions support distinguishable empirical commitments.

\textbf{Tier (iii): dynamics-internal grain selection.} Once the partition and intervention space are fixed, a further question remains about the grain at which the consciousness-relevant property is specified. The term dynamics-internal marks the fact that this grain is selected from the system’s own intervention-conditional transition structure, rather than from semantic labels, modelling convenience, or the theorist’s vocabulary. This idea recurs in the literature: integrated information theory’s \textit{exclusion postulate}, which selects the single grain at which the system’s integrated cause-effect structure is maximal; \textit{effective-information maximisation}, which selects the grain at which intervention-conditional transitions are most informative about successor states; and \textit{mechanistic individuation under intervention}, which selects components by their stable causal contributions under perturbation. In each case, the relevant grain is fixed by the system’s own intervention-conditional structure, in a way the theorist’s vocabulary cannot override. Different grains yield substantively different cause-effect structures, effective-information values, and mechanistic decompositions; these are real distinctions, not mere redescriptions. Given a fully specified system, intervention space, and criterion, observers should in principle converge on the same grain.

Tier (iii) observer-independence is conditional. It does not mean that a system selects its own boundary, interface, or intervention space without theoretical input. Those choices belong to tier (ii). The claim is instead that, once an empirically disciplined boundary, interface, and intervention space are fixed, the privileged grain or equivalence structure is determined by the system’s counterfactual transition organisation rather than by an observer’s semantic labelling. A canonical functional structure provides one example \citep{kanaima2026canonical}, which illustrates the fact that once interfaces are specified, states of the system can be partitioned according to possible future behaviour without relying on externally assigned labels.

Piccinini’s treatment of digital vehicles illustrates the same tier distinction in a concrete case. The labels assigned to digit types, such as ‘0’ and ‘1’, are arbitrary. But the grouping of physical microstates into digit types is not arbitrary when it is fixed by how the mechanism processes those states under the system’s operation. This is a case in which tier (i) labelling is observer-relative while the relevant individuation of computational vehicles is constrained by the mechanism’s own organisation.

Two kinds of claims should be kept apart at this point. The metaphysical claim is that a physical system has a causal-dynamical organisation in virtue of itself, fixed independently of any external labelling. The methodological claim is that doing science on such a system requires the partition and intervention-space choices described under tier (ii): coordinates, partitions, and labels have to be supplied before any structure can be investigated. The two are compatible. A system’s intervention-conditional structure is what it is whether or not a scientist is currently investigating it. An epistemic concern remains: reconstructing intrinsic causal structure from finite observations is often difficult. But this difficulty bears on what we can know about a system, not on what the system has. Cumulative theory-building, intervention across multiple modalities, and comparison of competing partitions are the standard scientific tools for narrowing the epistemic gap; the gap does not show that the structure being investigated is itself observer-imposed.

Pressing the objection through to tier (iii) requires the further premise that intervention-conditional identification at tiers (ii) and (iii) reduces to symbolic relabelling at tier (i). \citet[Sections 3.2--3.3]{lerchner2026abstraction} argues for this premise: Section 3.2 establishes the general claim that any identification of a physical state as a ‘readout’ or ‘hidden state’ already involves alphabetisation by a mapmaker, and Section 3.3 applies this directly to mechanistic individuation, extending by parity to intervention-based individuation. But the argument turns on conflating two operations: identifying a state by what it does under intervention is not the same as assigning it a label from a predefined alphabet. The two differ in their answerability to causal-structural facts. Different labellings of the same trajectory yield the same predictions under intervention, because relabelling preserves the dynamics; different intervention spaces or grain choices yield distinguishable predictions under intervention, because they partition the system’s causal structure differently. The Lerchner-style reply (that fixing an intervention space \textit{is} fixing an alphabet of admissible perturbations) collapses this distinction by treating both operations as equally arbitrary. They are not: a labelling cannot be empirically wrong about the system, only differently named, whereas an intervention-space or grain choice can be empirically wrong about the system, because it predicts effects that interventions then fail to produce. Label selection at tier (i) leaves intervention-conditional dynamics intact; grain selection at tier (iii) yields different decompositions of the system under intervention. The melody example, which illustrates the first kind of equivalence, does not by itself establish that the second kind is equally arbitrary.

The three-tier analysis applies beyond the mapmaker case. Syntax-is-not-semantics arguments and triviality arguments also turn on the claim that computational properties are observer-relative. The tier decomposition offers a common diagnostic resource: in each case, one must ask whether the objection targets tier (i) labelling specifically or extends to tier (iii) dynamics-internal individuation. The arguments are decisive against views that locate the relevant property at tier (i). Whether they extend to tier (iii) depends on whether the further premise (that all computational individuation reduces to labelling) can be sustained.

A remaining objection is that observer-relativity at tier (ii) might be carried over to tier (iii). Dynamics-internal grain selection does not introduce arbitrary label-selection of the tier (i) kind. It may still depend on the choice of a theoretical criterion, but that dependence is explanatory and empirically answerable, not stipulative. Once a partition and intervention space are given, the grain that maximises effective information (or that satisfies the alternative dynamics-internal criterion) is determined by the system’s transition structure. There is a fact of the matter at tier (iii) in a sense unavailable at tier (i). A residual observer-relativity enters at the level of selecting \textit{which} criterion to apply. The choice depends on what one takes consciousness to be. The dependence is theoretical-explanatory: predictions of competing theories of consciousness can be compared against neural correlates, behavioural indicators, and intervention-response patterns. A residual question is whether tier (iii) observer-independence is parasitic on the choice of partition at tier (ii). The position taken here is conditional: observer-independence at tier (iii) holds given an empirically disciplined choice of partition and intervention space at tier (ii), and intrinsic computational functionalism does not claim a substrate-internal selection principle for tier (ii). The asymmetry that carries the reply is that competing partitions support distinguishable intervention-conditional predictions, whereas competing labellings merely redescribe the same intervention-conditional structure.

\section{Applying the criteria to candidate architectures}

A contrast between two simple cases shows what the criteria filter. Consider a small Hopfield-style attractor network \citep{hopfield1982neural}, a recurrent network whose states settle into stable patterns (‘attractors’) stored by its weights, with symmetric weights and asynchronous updates (nodes updating one at a time), which together guarantee convergence to those attractors. Relabelling the nodes and correspondingly relabelling the weight matrix changes only the names, not the attractor landscape or transition structure. Properties preserved under such relabelling (basin structure, attractor patterns, convergence behaviour) are system-intrinsic in the sense (C1) requires: they are facts about the system that no relabelling creates or destroys.

Single-node perturbations that cross a basin boundary drive the system to a different stored pattern, producing distinguishable output distributions, while perturbations within a basin are corrected by the dynamics; the network’s intervention-conditional response thus exhibits the organised counterfactual structure (C2) requires, with the privileged grain identified by dynamics-internal measures such as effective information \citep{hoel2017map} or integrated cause-effect power \citep{albantakis2023integrated}. The attractor network satisfies (C1) and both sides of (C2) in a minimal formal sense: it has internal organisation with mutually constraining variables and a non-trivial intervention profile. It is not thereby a plausible candidate for consciousness, since it lacks the substantive conditions (recurrent self-modelling, global availability, temporal continuity) that consciousness theories additionally require.

Contrast a lookup-table system that reproduces the same input-output mapping by caching the network’s surface behaviour. Such a system may have internal physical states, but perturbing them does not reveal the basin structure, error-correction dynamics, or state-to-state dependencies of the attractor network. The table reproduces the input-output mapping without sharing the mechanism that generates it.

The system reproduces actual-input trajectories but lacks the integrated cause-effect structure that integrated information theory \citep{albantakis2023integrated} picks out, or the intervention-based causal informativeness that effective information \citep{hoel2017map} quantifies; it has no organised state space in which mutually constraining variables exhibit counterfactual signatures distinct from competing mechanisms. (C2) fails on the internal-organisation side because the lookup table does not share the target network’s organised state space at the relevant grain, in which mutually constraining variables exhibit the network’s characteristic counterfactual signatures; and on the counterfactual-structure side because the response to intervention on intermediate states is undefined or degenerate. The contrast shows the criteria are constructively discriminating: under fixed partition and intervention space, an attractor network supports the kind of intervention-conditional causal organisation (C2) requires, and a lookup-table replay does not, even when their surface behaviour matches.

A standard autoregressive transformer deployed with largely externalised conversational state provides a more realistic contrast. In standard transformer inference, hidden activations are transient; any persistence is typically carried by the context buffer, external memory, or model weights rather than by endogenous recurrent state dynamics \citep{kanai2025stream}. Whatever organisation the model exhibits during a single forward pass does not persist as an ongoing trajectory of mutually constraining variables. (C2)’s internal-organisation requirement asks whether the system has an organised state space in which variables sustain each other across time; a system that reinitialises at every session boundary does not meet this requirement at the session-to-session scale, regardless of the complexity of individual forward passes.

The harder cases are computational models of consciousness whose variables come with a rich theoretical vocabulary. As a test case, we consider Beautiful Loop, a recent active-inference theory of consciousness, asking how (C1) and (C2) apply to its proposed architecture without assessing whether the theory is correct \citep{laukkonen2025beautiful}.

Continuous belief variables update through gradient flow on a free-energy functional, an upper bound on surprise (the negative log model evidence) that the system minimises. Action and perception are coupled through a single generative model in which the body’s dynamics enter the inference loop as parameters that condition the system’s predictions \citep{friston2010free, tschantz2020learning}. Three jointly necessary conditions are proposed: a unified generative world model (an ‘epistemic field’); inferential competition that determines what enters that model; and epistemic depth, the recursive sharing of belief states across the system’s components \citep{laukkonen2025beautiful}. Beautiful Loop is directly targeted by Lerchner: he classifies it among the ‘recursive epistemic loops’ that reproduce the structural features of introspection without instantiating intrinsic meaning, charging the approach with the \textit{abstraction fallacy} \citep[Section 3.2]{lerchner2026abstraction}. A related but distinct objection, the \textit{transduction fallacy}, holds that adding sensors and actuators does not eliminate the observer, since an analogue-to-digital converter discretises voltages into integers under an interpretive scheme; Lerchner develops this for embodied/robotic systems generally \citep[Section 4.1]{lerchner2026abstraction}, and it applies to Beautiful Loop’s action-perception coupling by extension rather than by direct address. These three Beautiful Loop conditions provide useful test cases for the present criteria, though the mapping is inexact. The connection to (C1) remains uncertain.

Inferential competition and epistemic depth carry the kinds of organised structure and counterfactual signature (C2) requires, and admit direct intervention tests of the kind sketched below.\footnote{Any active-inference reading must engage three challenges. Predictive processing has been argued to fail as a unifying theory across the dimensions of generality, simplicity, homogeneity, and systematicity, and to invite a \textit{consistency fallacy} in which evidence merely consistent with a model is treated as confirmatory \citep{litwin2020unification}. The mortal-computation proposal associated with \citet{hinton2022forward} challenges the standard separation between algorithm and hardware by suggesting that some computations may be tied to particular physical hardware instances; in consciousness debates, this has been used to question substrate-independent digital implementation. The dispute narrows once carbon chauvinism is rejected on both sides and becomes one of degree about what physical features are required. A no-go theorem argues that, if consciousness is dynamically relevant, digital chips, engineered and validated so that their dynamics suppress any departure from a fixed computational program, cannot host the dynamics consciousness requires \citep{kleiner2024case}; whether register-level physical dynamics under intervention satisfy (C2) is not settled here and is treated as a general substrate matter in Section 10.}

A general worry about Bayesian-inferential architectures deserves direct treatment at this point. Active-inference architectures treat the system as inferring hidden causes of sensory inputs by minimising free energy, with the inferred posterior called the ‘recognition density’ and the system’s internal model called the ‘generative model’. A generative model is partly chosen by the modeller, and belief variables are parameters of that model. The worry is that the relevant property is then partly observer-relative at the level of the model, even if the underlying register dynamics are fixed by physics. Two responses are available. First, (C1) is satisfied at any level where the architecture’s dynamics are fixed by its own structure under intervention: if the recognition density updates through gradient flow whose intervention-conditional structure is the same across choices of variable names, the dynamical property is system-intrinsic in (C1)’s sense. Second, (C1)-satisfaction at the model level requires showing that the model-grain itself tracks system facts, in a way the modeller’s convenience cannot dictate, by some dynamics-internal criterion. Whether a given active-inference implementation satisfies this stronger requirement is an empirical question, not one that can be settled from the architecture’s vocabulary alone.

(C1) is the harder case. At the physical implementation level, the relevant question is whether the register dynamics that realise the update (voltage transitions, charge redistributions, gate switching) are fixed under intervention independently of variable naming. At the belief-variable level, (C1) is satisfied only if the belief-variable description is itself fixed by the system’s intrinsic dynamics rather than chosen by the modeller, and whether any active-inference implementation satisfies that stronger requirement is currently under empirical investigation. The correspondence is therefore conceptual only. Given the circuit-level partition and intervention space, the subsequent transition structure is not created by semantic labelling: voltage transitions and gate switching are physical events that no relabelling alters. But Beautiful Loop as a consciousness theory operates at the belief-variable level, and (C1)-satisfaction there requires showing that the belief-variable description tracks system facts about the dynamics independently of the modeller’s choice. This is the harder and still-open question. Lerchner’s transduction-fallacy objection, that an analogue-to-digital converter discretises voltages under an interpretive scheme, is best located within the tier framework. The discretisation belongs to tier (ii): it is an interface choice that conditions the analysis, whereas a tier (i) label is imposed after the fact and leaves the dynamics intact. The register dynamics that result from this interface are tier (iii) facts about the system; they belong to the system’s own causal structure regardless of the observer’s vocabulary. Biological neural systems also involve forms of discretisation of their own: action potentials are all-or-nothing events, and synaptic transmission is quantal. Lerchner himself distinguishes thermodynamic discretisation from the semantic alphabetisation he ultimately objects to \citep[Section 2.4]{lerchner2026abstraction}, so this parallel is a limited one. The narrower point is that any disqualification grounded in discretisation alone, considered apart from semantic assignment, would apply symmetrically to biological and digital substrates.

The architecture admits direct counterfactual tests of the kind (C2)’s second operational test requires. Intervening on prior preferences shifts policy selection \citep{tschantz2020learning}; analogous interventions on the recognition density and action policies follow from the same framework. If the belief-variable level turns out to maximise a suitable dynamics-internal criterion, such as effective information, it would be a candidate privileged grain for applying (C1) and (C2). On a digital implementation, continuous belief variables are represented by floating-point encodings, which are conventional at the algorithmic level. But the register dynamics that realise the computation are not created by those conventions. The relevant question is therefore not merely whether, given a circuit-level partition, the substrate dynamics proceed independently of semantic labelling; rather, it is whether the belief-variable grain is itself system-intrinsic.

The mapping is testable in the sense that one can specify what would defeat it. A world model sustained only by external readout would fail (C1). Inferential competition stipulated by an external scheduler, not emerging from coupled belief dynamics, would defeat (C2) on its internal-organisation side. A recorded-trace implementation reproducing the same surface dynamics under flat responses to intervention would defeat (C2) on its counterfactual-structure side. Whether Beautiful Loop satisfies (C1) and (C2) is empirically open. The value of the example is that the criteria specify failure conditions: what would have to be absent for a system to fail, and what would have to be shown for it to pass. The example serves to make the criteria’s demands concrete, not to establish that a satisfier exists.

\section{Partition selection and empirical constraint}

A natural objection to the Section 7 concession at tier (ii) is that partition selection, because it is theory-driven, smuggles observer-relativity back into the position: empirical constraint just relocates the observer’s choice to the choice of theory. The reply is that theory-driven revision of partitions is empirically disciplined rather than stipulative. This does not mean that data uniquely determine a partition. It means that partitions are answerable to empirical and explanatory constraints in a way arbitrary labels are not, and this is part of how consciousness science already proceeds. Theories of consciousness make distinguishable predictions under their partition commitments, and a partition that conflicts with correlational evidence, intervention responses, or theory-derived indicators is empirically inadequate rather than merely a different observer’s choice.

The indicator-property approach derives diagnostic markers from neuroscientific theories and assesses whether candidate systems satisfy them (\citealp{butlin2023consciousness}; \citealp{butlin2025identifying}). This approach illustrates the point because an indicator cannot be applied without deciding which components, processes, and causal relations in the target system are the relevant ones, and it is the assignments that yield predictions surviving empirical test that constrain the partition. Prioritising indicators that cannot easily be designed without also creating consciousness \citep{butlin2025identifying} reflects a parallel concern: that the relevant property should not be surface-reproducible without the underlying organisation.

A consciousness-research example helps locate the disagreement. Gross anatomical distinctions between posterior and prefrontal cortex are not themselves created by consciousness theory, even if finer parcellations may vary with method. Competing theories disagree about which anatomically individuated regions carry the consciousness-relevant computational structure, not about anatomy itself. The disagreement yields distinguishable predictions about correlates of conscious experience under conditions where conscious content varies, and these predictions are tested against intervention data. Finer-grained partition choices (single neurons versus populations, membrane potentials versus firing rates) face similar constraints, though the empirical discipline is harder to apply and the underdetermination more pronounced.

The partition difference is at the level of theoretical commitment about which physically individuated systems are consciousness-relevant, and that commitment is distinguishable under intervention. The tier (ii) cost is real but tractable: partition choices are observer-involving, yet empirically revisable in a way arbitrary labels are not.

\section{Implications for current artificial systems}

Intrinsic computational functionalism has cautious implications for artificial consciousness research. The criteria do not decide between biological naturalism and substrate-general functionalism; they state the structural burden that either side must meet.

Behavioural and linguistic criteria are vulnerable to anthropomorphic overinterpretation, so current AI systems should not be assumed conscious merely because they display sophisticated language, reasoning, or self-description \citep[cf.][]{chalmers2023could}. The absence of biological tissue does not by itself settle the question, since if consciousness depends on intrinsic organisation that can be realised in non-biological substrates, artificial consciousness remains possible in principle. Digital implementation is neither automatically disqualifying nor automatically sufficient. What matters is whether digital or hybrid systems can instantiate the intrinsic structural properties that (C1) and (C2) require at the consciousness-relevant grain. AI consciousness research should attend to recurrence, integration, counterfactual response under intervention, self-modelling, and embodiment. AI welfare debates \citep{long2024taking} should treat verbal reports as insufficient by themselves, and should ask whether such reports are connected to internal functional organisation of the relevant kind (\citealp{butlin2023consciousness}; \citealp{butlin2025identifying}).

The role-realiser distinction sharpens the substrate question. Block frames it as a contrast between the biological realiser (‘meat’) and the computational role, arguing that introspection alone does not settle which is consciousness-crucial \citep{block2026can}.

(C1) and (C2) constrain the kind of physical organisation a realiser must have, with (C2) placing the most direct constraints on temporal coupling and intervention-responsiveness. If a theory requires temporally extended causal coupling, a substrate whose relevant coupling is too transient to support that organisation fails the theory’s realiser condition, regardless of input-output behaviour.

(C2)’s counterfactual-structure side is satisfied where a substrate’s components respond to intervention with stable, characteristic effect distributions, and fails where substrates produce degenerate responses to intervention on intermediate states even when they reproduce surface dynamics. Whether biological substrates have implementation features no non-biological substrate can replicate is an empirical matter, and neuromorphic and analogue substrates should be evaluated on their own terms rather than accepted or rejected by association with clocked digital implementation.

The simulation-instantiation distinction is sharpened by these criteria but remains unsettled. Consider the chloroplast case. A simulation of photosynthesis does not produce glucose because glucose production requires the chemical organisation that actual chloroplasts have, which the GPU running the simulation does not share. The components of a GPU running a simulation do not respond to interventions on substrate chemistry, light, or temperature with the effect distributions characteristic of chloroplast dynamics. But the chloroplast case concerns a specific demand: that the actual chemical mechanisms be present, whereas (C2)’s structural-conditional counterfactuals concern the intervention-conditional relations among the system’s variables, without requiring that those variables produce the further metaphysical effects the simulated kind would have.

The relevant disanalogy is that glucose production demands the chemistry: without the relevant chemical mechanisms, glucose is not produced. Consciousness, by contrast, may or may not demand the underlying chemistry, depending on which level the consciousness-relevant property lives at. The chloroplast case thus does not, by itself, settle the AI case. If the consciousness-relevant property lives at the molecular-biochemical level, digital substrates fail for the same reason simulated chloroplasts do. If it lives at a coarse-grained dynamical level characterisable by integrated cause-effect structure under intervention, the question becomes whether a digital system realises such organisation at the right grain. The criteria specify what any candidate substrate must satisfy at whichever level turns out to matter, without committing to a level. In this sense, (C1) and (C2) leave the biological-naturalism versus substrate-general functionalism dispute open: they describe the structural burden either side must shoulder.

\section{A unified research orientation}

We propose intrinsic computational functionalism as a shared methodological constraint on computational approaches to consciousness.

The convergent programmes may ultimately disagree on details, but they share the methodological core: consciousness cannot be grounded in arbitrary external interpretation, physical realisation is required, and the focus is internal causal organisation. (C1) and (C2) specify necessary preconditions that any computational approach to consciousness must satisfy if it is to avoid observer-relativity.

What the criteria deliver is a constraint set. The constraint is formally specifiable: for any system whose transition structure and intervention space can be written down, (C1) and (C2) pose well-defined questions evaluable without recourse to brain-specific anatomy, in line with the requirement that consciousness theories apply across substrates without dependence on external interpretation. Whether those questions yield decisive verdicts in practice depends on the empirical and theoretical resources available.

The paper has not settled which intrinsic structures are consciousness-relevant, nor whether digital substrates can realise the relevant organisation at the consciousness-relevant grain. That the consciousness-relevant level is individuable by intervention-conditional internal structure alone remains a substantive commitment rather than a demonstrated conclusion. What the paper has argued is a diagnostic: anti-computational arguments succeed against properties located at the level of imposed labels, but their extension to dynamics-internal organisation requires a further premise that has not been supplied. The criteria and the tier decomposition make this diagnostic precise.

\citet{borges1998exactitude}, in \textit{On Exactitude in Science}, imagined a map drawn to the scale of the empire it depicted, abandoned to the desert when its uselessness became clear. Once a theory of consciousness fixes which features of the territory matter, the carving that the system supports under intervention is no longer the cartographer’s invention. Whether any current theory has fixed those features correctly remains the open question, and which intrinsic structures make a system a subject of experience is the question consciousness science needs to answer.

\section*{Declarations}

\textbf{Funding}: This work was supported by Fudan University and the Pivotal Fellowship Fund.

\textbf{Competing interests}: The authors declare no competing interests.

\textbf{Author contributions}: R.K.\ initiated the project. S.M.\ and R.K.\ contributed equally to this work. Both authors read and approved the final manuscript.

\setlength{\bibsep}{8pt}
\renewcommand{\bibitemsep}{0pt}
\bibliography{references}

\end{document}